\documentclass[aps
, prl
, twocolumn
, nofootinbib
, preprintnumbers
]{revtex4-1}

\usepackage{amsmath,amssymb}
\usepackage{graphicx}
\usepackage{epstopdf}
\usepackage{dcolumn}% Align table columns on decimal point
\usepackage{bm}% bold math
\usepackage{color}
\usepackage{comment}
\usepackage{verbatim}
\usepackage[utf8]{inputenc}
\usepackage[colorlinks=true
,urlcolor=blue
,anchorcolor=blue
,citecolor=blue
,filecolor=blue
,linkcolor=blue
,menucolor=blue
,pagecolor=blue
,linktocpage=true
,pdfproducer=medialab
,pdfa=true
]{hyperref}

\begin{document}

%%%%%%%%%%%%%%%%%%%%%%%%%%%%%%%%%%%%%%%%%%%

\def\dd{\mathrm{d}}
\def\mcH{\mathcal{H}}
\def\mcP{\mathcal{P}}
\def\mcR{\mathcal{R}}
\def\tot{{\rm tot}}
\def\obs{{\rm obs}}
\def\mi{{\rm min}}
\def\ma{{\rm max}}
\def\Mpl{M_{\rm Pl}}
\def\GeV{{\rm GeV}}
\def\CMB{{\rm CMB}}
\def\f{{\rm f}}
\def\G{{\rm G}}
\def\reh{{\rm reh}}
\newcommand{\Red}[1]{\textcolor{red}{\sffamily #1}}
\newcommand{\Blue}[1]{\textcolor{blue}{\sffamily #1}}
\newcommand{\Mag}[1]{\textcolor{magenta}{\sffamily #1}}
\newcommand{\mathred}[1]{\textcolor{red}{#1}}
\newcommand{\bae}[1]{\begin{align} #1 \end{align}}
\newcommand{\ee}{\mathrm{e}}
\newcommand{\calP}{\mathcal{P}}
\newcommand{\ns}{n_{{}_\mathrm{S}}}

\def\Hinf{H_{\rm inf}}
%%%%%%%%%%%%%%%%%%%%%%%%%%%%%%%%%%%%%%%%%%%%%%%%%%%%%%%%%%%%%%%

\title{
Does the detection of primordial gravitational waves exclude low energy inflation?
}

\author{Tomohiro Fujita}
\affiliation{Stanford Institute for Theoretical Physics, Stanford University, Stanford, CA 94306, U.S.A.}
\affiliation{Department of Physics, Kyoto University, Kyoto, 606-8502, Japan}
\author{Ryo Namba}
\affiliation{Department of Physics, McGill University, Montr\'{e}al, QC, H3A 2T8, Canada}
\author{Yuichiro Tada}
\affiliation{Institut d’Astrophysique de Paris, UMR-7095 du CNRS, Université Pierre et Marie Curie, 98 bis bd Arago, 75014 Paris, France}
\affiliation{Sorbonne Universités, Institut Lagrange de Paris, 98 bis bd Arago, 75014 Paris, France}
\affiliation{Kavli Institute for the Physics and Mathematics of the Universe (WPI), UTIAS, The University of Tokyo, Kashiwa, Chiba 277-8583, Japan}
\affiliation{Institute for Cosmic Ray Research, The University of Tokyo, Kashiwa, Chiba 277-8582, Japan}

\begin{abstract}
We show that a detectable tensor-to-scalar ratio $(r\ge 10^{-3})$ on the CMB scale can be generated even during extremely low energy inflation which saturates the BBN bound $\rho_{\rm inf}\approx (30 {\rm MeV})^4$. The source of the gravitational waves is not quantum fluctuations of graviton but those of $SU(2)$ gauge fields, energetically supported by coupled axion fields. The curvature perturbation, the backreaction effect and the validity of perturbative treatment are carefully checked. Our result indicates that measuring $r$ alone does not immediately fix the inflationary energy scale.
\end{abstract}
\maketitle
\preprint{IPMU 17-0073}
%%%%%%%%%%%%%%%%%%%%%%%%%%%%%%%%%%%%%%%%%%%%%%%%%%%%%%%%%%%%%%%

%************************************************************************************%
%
%
%
%====================================================================================%
\section{I. Introduction}
%====================================================================================%

The inflationary paradigm has been successful over the past few decades to serve as a mechanism to produce the observed inhomogeneities in the universe such as the cosmic microwave background (CMB) anisotropies and large-scale structure (LSS), while resolving the conceptual difficulties in the hot big bang scenario. An important prediction in the framework is generation of the B-mode polarization in the CMB 
\cite{Kamionkowski:1996zd, *Seljak:1996gy}, 
whose signal is conventionally quantified by the tensor-to-scalar ratio $r\equiv \mcP_h/\mcP_\zeta \vert_{k = k_\CMB}$. 
The current bound is $r < 0.07$ at $k_{\rm CMB} = 0.05 \, {\rm Mpc}^{-1}$ with $95\%$ confidence \cite{Array:2015xqh}, and a number of proposed missions are expected to improve the bound to ${\cal O}(10^{-3})$ (see e.g.~\cite{Abazajian:2013vfg}).
The conventional relationship between the tensor-to-scalar ratio and the Hubble parameter during inflation is
\begin{equation}
r_{\rm vac} = \mcP_\zeta^{-1}\frac{2\Hinf^2}{\pi^2 \Mpl^2} \approx 10^{-3} \left(\frac{\Hinf}{8\times 10^{12}\GeV}\right)^2 \; ,
\label{vac r}
\end{equation}
where $H_{\rm inf}$ is the Hubble parameter during inflation and $\mcP_\zeta \approx 2.2 \times 10^{-9}$ has been used \cite{Ade:2015lrj}. An immediate implication of \eqref{vac r} is that detection of $r$ would fix the inflationary scale at such high energy levels as beyond our current experimental reach.

Considering the ongoing and upcoming experimental efforts for B-mode detection, it is right time to test the validity of the conventional prediction \eqref{vac r}. 
In general, the value of $r$ at cosmological scales can be estimated as the spectrum of the energy fraction of gravitational wave (GW) at the horizon crossing divided by $\mcP_\zeta$
\begin{equation}
r \simeq \mcP_\zeta^{-1} \frac{1}{\rho_{\rm inf}} \frac{\dd \rho_{\rm GW}}{\dd \ln k}\Big|_{k= a \Hinf},
\end{equation}
where $\rho_{\rm inf}\equiv 3\Mpl^2\Hinf^2$ and $\dd \rho_{\rm GW}/\dd \ln k \simeq H^2\Mpl^2 \mcP_h$
at the horizon crossing. 
The energy density of GW from the vacuum fluctuations produced during the quasi de Sitter expansion must be characterized by the Hubble scale $\dd \rho_{\rm GW}^{\rm vac}/\dd \ln k \simeq \Hinf^4$, leading to the conventional relation $r_{\rm vac}\propto \Hinf^2$. 

On the other hand, if GW is induced by another energy source, the conventional relation \eqref{vac r} may be altered. Provided that an energy source $\rho_s$ generates GWs with efficiency $\gamma$, one generally expects 
\begin{equation}
r \simeq \mcP_\zeta^{-1}\frac{\gamma}{\rho_{\rm inf}} \frac{\dd \rho_s}{\dd \ln k}\Big|_{k = a \Hinf},
\label{r-general}
\end{equation}
which can be significant even if $\rho_s\ll\rho_\text{inf}$ and $\gamma\ll1$ thanks to the smallness of $\mathcal{P}_\zeta$. Conventionally, however,  an efficient energy transfer from a source to GW has been assumed to be rather difficult. The reasoning is rooted in the {\it decomposition theorem} in cosmology, which states that perturbations around a homogeneous and isotropic background can be decomposed into scalar, vector and tensor sectors that are mutually decoupled at the linearized order.
Since GW is the only tensor degree of freedom in the Einstein gravity, 
we have no choice but use the source term from scalar $\delta S$ or vector perturbation $\delta V_i$ which is schematically written as
\begin{equation}
\Box h_{ij} (t,\bm x)= O_{ij}^{(S)}(t, \bm \partial) \, \delta S(t,\bm x) +  O_{ijk}^{(V)}(t, \bm\partial) \, \delta V_k(t,\bm x) \; ,
\label{Box-h}
\end{equation}
where $O_{ij}^{(S)}$ and $O_{ijk}^{(V)}$ are operators traceless and transverse in the indices $ij$ that depend on time and spatial derivatives.
However, the decomposition theorem bans
the existence of such operators at the linear order.
Although the second order effects (e.g.~$\partial_i \delta S \partial_j \delta S,\ \delta V_i \delta V_j$) are allowed to generate GW, the efficiency of the energy transfer is suppressed, because the coefficients of the source term effectively becomes the order of perturbation, $O_{ij}^{(S)}, O_{ijk}^{(V)} = \mathcal{O}(\delta S,\delta V_j)$ \cite{Cook:2011hg, *Senatore:2011sp, *Barnaby:2012xt,*Biagetti:2013kwa, *Biagetti:2014asa, *Fujita:2014oba, *Mirbabayi:2014jqa}.

There is a loophole in this argument. If $O_{ijk}^{(V)}$ in \eqref{Box-h} consists of the background vector field $\bar{V}_i(t)$, GW can be sourced at linear order by $\bar{V}_i\delta V_j$. 
It is known that $SU(2)$ gauge fields can
achieve this without disrupting background isotropy by taking
a particular configuration.\footnote{This does not restrict possible models to those with $SU(2)$ only, 
as long as the symmetry in the models allow this configuration.}
Moreover this isotropic configuration is realized as an attractor solution, if $SU(2)$ gauge fields are coupled to a rolling pseudo-scalar field 
\cite{Adshead:2012kp, *Adshead:2013nka, *Maleknejad:2013npa}.
Therefore $SU(2)$ gauge fields can source the GW through the terms $\bar{V}_i\delta V_j$ without violating the isotropy of the universe at the linear order, thus with a high efficiency of the energy transfer.

As we shall see later, the energy source $\rho_s$ to generate GW is the (linear) perturbation of an $SU(2)$ gauge field. It is produced as quantum fluctuations and thus acquires the amplitude $\mathcal{O}( \Hinf)$ around the horizon crossing. In addition, however, it experiences a transient instability around horizon crossing and is amplified by an exponential factor. As a result, the energy fraction of the source and the efficiency factor of energy transfer in \eqref{r-general} are given by
\begin{equation}
\frac{1}{\rho_{\rm inf}}\frac{\dd \rho_s}{\dd \ln k} \sim \frac{\Hinf^2}{\Mpl^2}e^{4m_Q},
\quad
\gamma \sim \frac{\rho_A}{\rho_{\rm inf}}\equiv \Omega_A,
\label{rhogamma}
\end{equation}
where $s$ now denotes the perturbation of $SU(2)$ gauge field, $\rho_A$ is its background energy density, 
and $m_Q$ is the $SU(2)$ mass parameter in the units of $\Hinf$. 
For values of $m_Q$ with $\Hinf \sqrt{\Omega_A} \, e^{2m_Q} \gtrsim {\cal O}(10^{12}) \, \GeV$, one can realize a detectable $r$ even in the case of low-energy inflation.

%====================================================================================%
\section{II. Spectator axion-$SU(2)$ Model}
%====================================================================================%

In our consideration of GW production, we leave the gravity sector as the standard Einstein-Hilbert and the inflation model unspecified, which is also responsible for generating the observed curvature perturbation. We then consider the axion-$SU(2)$ sector with the action \cite{Dimastrogiovanni:2016fuu} (see also \cite{Dimastrogiovanni:2012st}):
\begin{multline}
\mathcal{L}_{\chi A} =
- \frac{1}{2}(\partial_\mu \chi)^2 
- V(\chi)
-\frac{1}{4} F^a_{\mu\nu}F^{a\mu\nu} +\frac{\lambda}{4f}\chi F_{\mu\nu}^a \tilde{F}^{a\mu\nu},
\label{lagrangian}
\end{multline}
where $\chi$ is a pseudo-scalar field (axion) with a cosine-type potential $V(\chi) = \mu^4 \left[1+\cos (\chi/f)\right]$ with dimensionful parameters $\mu$ and $f$, $F^a_{\mu\nu}\equiv 2 \partial_{[\mu} A_{\nu]}^a - g \epsilon^{abc} A^b_\mu A^c_\nu$ and $\tilde{F}^{a\mu\nu}$ are the field strength of $SU(2)$ gauge field and its dual, respectively, and $\lambda$ is a dimensionless coupling constant.

At the background level, it is shown that the isotropic configuration of the $SU(2)$ gauge fields, $A^a_0=0$ and $A_i^a = \delta^a_i a(t) Q(t)$, is an attractor solution while the vev of $\chi(t)$ slowly rolls down its potential 
\cite{Adshead:2012kp, *Adshead:2013nka, *Maleknejad:2013npa, Dimastrogiovanni:2016fuu}. 
At the perturbation level, $\delta A_\mu^a$ contains two scalar $\delta Q$, $M$, two vector $M_i$ and two tensor $t_{ij}$ polarizations as dynamical degrees of freedom \cite{Adshead:2012kp, *Adshead:2013nka, *Maleknejad:2013npa, Dimastrogiovanni:2016fuu}. Interestingly, $t_{ij}$ is coupled to the metric tensor modes $h_{ij}$ already at the linear order, and only one circular polarization mode of $t_{ij}$ is substantially amplified due to a transient instability around the horizon crossing. It then efficiently sources one polarization of GW $h_{ij}$  at the linear order, if $m_Q\equiv gQ/H>\sqrt{2}$~\cite{Dimastrogiovanni:2012ew}. Therefore we focus on $t_{ij}$ among the perturbations of $A_\mu^a$.

The Einstein equation at the background yields
\begin{align}
3\Mpl^2 H^2 &= \rho_\phi +\rho_\chi +\rho_A+\rho_t,
\label{Friedmann}
\\
-\dot{H}/H^2 &= \epsilon_\phi +\epsilon_\chi +\epsilon_A +\epsilon_t,
\label{Friedmann2}
\end{align}
where $\rho_\chi=\dot{\chi}^2/2+
V(\chi)$, $\rho_A=3 \epsilon_A \Mpl^2 H^2 / 2$, $\epsilon_A=\epsilon_E+\epsilon_B, \ \epsilon_E\equiv (\dot{Q}+HQ)^2/\Mpl^2H^2,\ \epsilon_B\equiv g^2Q^4/\Mpl^2H^2$,  $\epsilon_\chi=\dot{\chi}^2/2\Mpl^2H^2$, and dot denotes the cosmic time derivative.
The inflaton part $\rho_\phi$ and $\epsilon_\phi\equiv -\dot{\rho}_\phi/6\Mpl^2 H^3$ depend on the inflation model, and $\rho_t$ and $\epsilon_t\equiv -\dot{\rho}_t/6\Mpl^2 H^2$ denote the contributions from the perturbation $t_{ij}$ on the background dynamics, which will be discussed later. The equations of motion for $\chi(t)$ and $Q(t)$ are
\begin{align}
& \ddot{\chi}+3H\dot{\chi}-\frac{\mu^4}{f}\sin \left(\frac{\chi}{f}\right)
+\frac{3g\lambda}{f}Q^2 \left(\dot{Q}+HQ\right)+\mathcal{T}^\chi_{BR}=0,
\label{chiEoM}
\\
& \ddot{Q}+3H\dot{Q} +\left(\dot{H}+2H^2\right)Q +2g^2 Q^3 -\frac{g\lambda}{f} Q^2 \dot{\chi}+\mathcal{T}_{BR}^Q=0,
\label{QEoM}
\end{align}
where we include the backreaction terms, $\mathcal{T}_{BR}^Q$ and $\mathcal{T}^\chi_{BR}$, from $t_{ij}$. Without the backreaction, one can show that the effective potential of $Q$ uplifted by the coupling to $\chi$ acquires a non-zero minimum at $Q_{\min}\simeq \left(\mu^4 \sin(\chi/f)/3g\lambda H\right)^{1/3}$,
if $\chi$ slowly rolls and the coupling is sufficiently strong \cite{Adshead:2012kp, *Adshead:2013nka, *Maleknejad:2013npa, Dimastrogiovanni:2016fuu}.

The tensor perturbations consist of $t_{ij}$ and $h_{ij}$, and each of them can be decomposed into the circular polarization modes $t_{R/L}$ and $h_{R/L}$, respectively. At the linearized order, one finds their equations of motion coupled together among the same polarizations, written in the Fourier space as \cite{Dimastrogiovanni:2016fuu},
\begin{align}
&\partial_x^2 t_{R,L}+\left[1+\frac{2m_Q \xi}{x^2}\mp2\frac{m_Q+\xi}{x}\right]t_{R,L} \approx 0
%\mathcal{S}^t_{R,L}, 
\label{tEoM}\\
&\partial_x^2 \psi_{R,L}+\left(1-\frac{2}{x^2}\right)\psi_{R,L} \approx \mathcal{S}^\psi_{R,L},
\label{psiEoM}
\end{align}
where $x\equiv k/aH$ and  $\psi_{R,L}(t,k)$ are the mode functions of the canonical gravitational wave, $\psi_{ij} \equiv a\Mpl h_{ij}/2$. 
While $t_{R/L}$ are sourced by $\psi_{R/L}$ in principle, the former is always parametrically larger than the latter for our concern, and thus ignoring the right-hand side of \eqref{tEoM} is a justified approximation.
We have also neglected slow-roll suppressed and subdominant terms in \eqref{tEoM} and \eqref{psiEoM}. Here, $\xi(t)\equiv \lambda \dot{\chi}/2fH$ is well approximated by $m_Q+m_Q^{-1}$ in the slow-roll regime.
Without loss of generality $m_Q$ is assumed to be positive, and then $t_R$ becomes unstable for $x_{\rm max}>x>x_{\min}$,
with $x_{\max,\min}\equiv m_Q+\xi\pm(m_Q^2+\xi^2)^{1/2}$. 
Assuming $m_Q=\text{const.}$, we obtain the homogeneous solution to \eqref{tEoM} as
\begin{equation}
t_R(t,k)=\frac{1}{\sqrt{2k}} \, e^{\frac{\pi}{2}(m_Q+\xi)} \, W_{\beta,\alpha}\left(-\frac{2ik}{aH}\right),
\label{homogeneous solution}
\end{equation}
where $W_{\beta,\alpha}(z)$ is the Whittaker function with $\alpha \equiv -i\sqrt{2m_Q \xi -1/4}$ and $\beta \equiv -i(m_Q+\xi)$.
We have used the WKB solution in the sub-horizon limit, $t_R(k/aH\to \infty)=(2k)^{-1/2} (2x)^\beta e^{i x}$, as the initial condition.
Then $t_R$ is amplified around the horizon crossing by the factor of $e^{\frac{\pi}{2}(m_Q+\xi)} W_{\beta,\alpha}(-2i x_{\min})\approx e^{1.85m_Q}$, while it decays as matter, $\rho_t \propto a^{-3}$ i.e.~$t_{R}\propto a^{-1/2}$, on super-horizon scales.
The source term for $\psi_{R/L}$ reads
\begin{align}
\mathcal{S}^\psi_{R,L}&\equiv
\frac{2\sqrt{\epsilon_{E}}}{x}\partial_x t_{R,L}+\frac{2\sqrt{\epsilon_{B}}}{x^2}\left(m_{Q}\mp x\right)t_{R,L}\,,
\label{tsourceterm}
\end{align}
and the generated $t_R$ sources $\psi_R$, producing additional GW.
Using \eqref{homogeneous solution}, one can obtain the sourced $\psi_R$ by using Green's function method, giving the GW power spectrum
\begin{equation}
\mcP_h^{(s)} = \frac{\epsilon_B H^2}{\pi^2 \Mpl^2}\mathcal{F}^2(m_Q),
\label{Phs}
\end{equation}
where $\mathcal{F}^2\approx 2e^{3.62m_Q}$ and its full expression can be found in \cite{Dimastrogiovanni:2016fuu}. Note that \eqref{homogeneous solution} and \eqref{Phs}
assume constant $\epsilon_B$, $m_Q$ and $\xi$, while to determine their values and time variations one needs to solve the background dynamics, \eqref{Friedmann}--\eqref{QEoM}.

%====================================================================================%
\section{III. Checklist}
%====================================================================================%

In order to settle the final allowed strength of GW signals from this model,
we need to ensure some computational and observational consistencies. We list them and show the resulting parameter region in the following subsections.

%
%///////////////////////////////////////////////////////////////////////////////////%
\begin{figure}[tbp]
\includegraphics[width=\hsize]{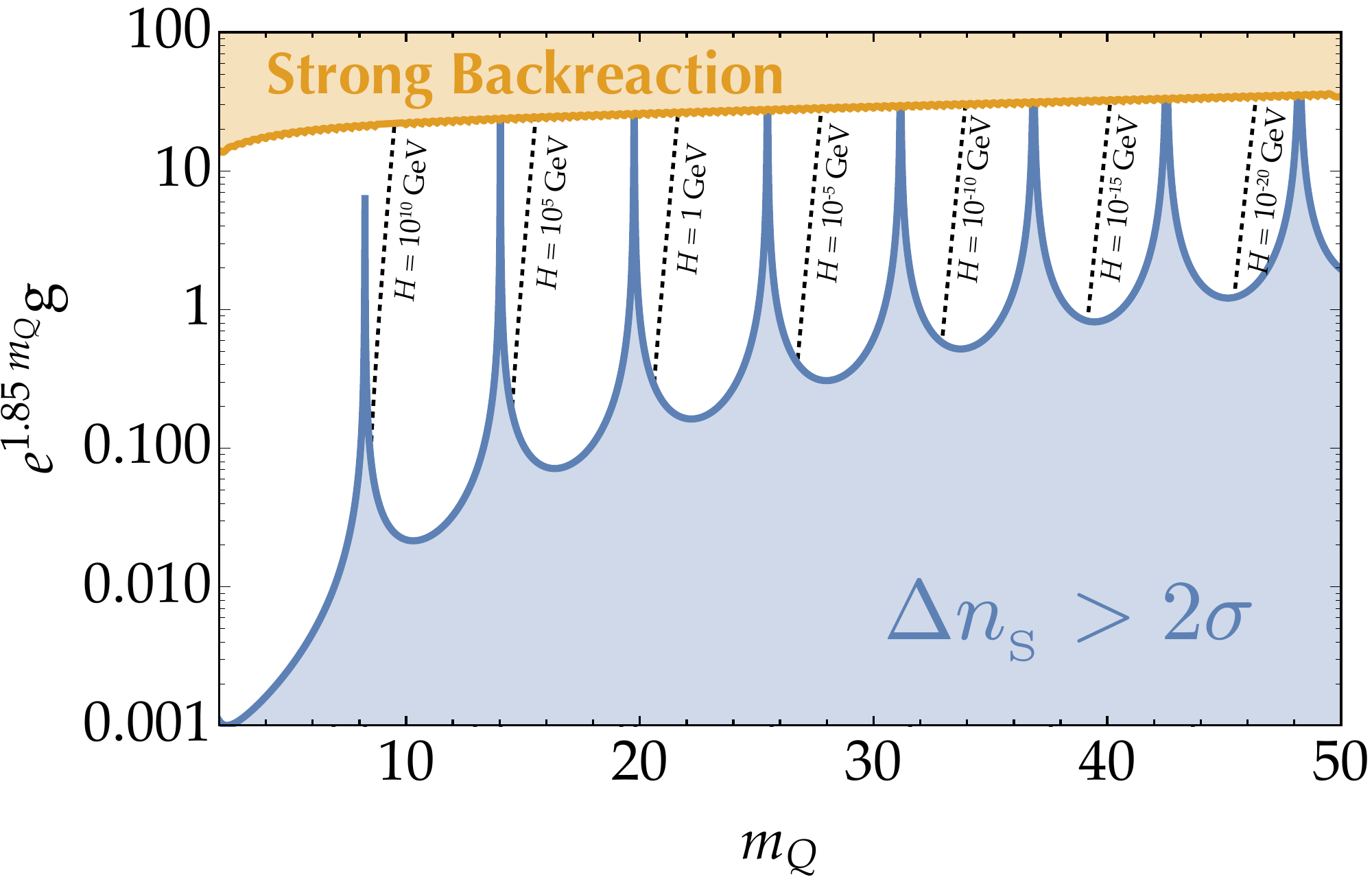}
\caption
{The allowed values of the $SU(2)$ gauge self-coupling constant $g$. 
Since this constraints are proportional to $e^{1.85m_Q}$ as mentioned in the main text, the coupling constant $g$ shown in the plot is rescaled by this factor.
In the upper yellow shaded region, 
the backreaction is expected to be strong and disrupts the background evolution.
In the lower blue shaded region, the energy fraction of the gauge field is significant enough to make the scalar spectral index becomes too red beyond the 2$\sigma$ region of Planck constraints for $r=10^{-3}$.
The black dotted contours for the values of $\Hinf$ are superimposed in the case with $r=10^{-3}$.}
\label{fig:checks}
\end{figure}
%///////////////////////////////////////////////////////////////////////////////////%
%

%====================================================================================%
\subsection{A. Backreaction}
%====================================================================================%

The produced $t_R$ \eqref{homogeneous solution} backreacts on the background dynamics through eqs.~\eqref{Friedmann}--\eqref{QEoM} with the terms
\begin{align}
\rho_t &=\frac{1}{2a^2}\int\frac{\dd^3 k}{(2\pi)^3} \left[ \left|\dot{t}_R \right|^2+ \left(\frac{k^2}{a^2} - 2m_Q H\frac{k}{a}\right)\left|t_R \right|^2\right],
\label{rhot}
\\
\mathcal{T}_{BR}^\chi &= -\frac{\lambda}{2a^3 f} \frac{\dd}{\dd t}   \int \frac{\dd^3 k}{(2\pi)^3} \left(a m_Q H-k\right)|t_R|^2,
\label{TBRchi} \\
\mathcal{T}_{BR}^Q &=\frac{g}{3a^2}   \int \frac{\dd^3 k}{(2\pi)^3} \left(\xi H-\frac{k}{a}\right)|t_R|^2,
\label{TBRQ}
\end{align}
where we ignore the sub-leading backreaction from $t_L$ or $\psi_{R,L}$.
We first estimate these contributions analytically. Using \eqref{homogeneous solution} and background relation $\xi \cong m_Q + m_Q^{-1}$ and changing variables into $x = k/aH$ with the integration domain $0 < x < x_{\rm max}$, 
one can write $|\rho_t| = H^4 {\cal I}_\rho(m_Q)$, $|{\cal T}_{BR}^\chi| = \lambda H^4 {\cal I}_\chi(m_Q) / f $ and ${\cal T}_{BR}^Q = g H^3 {\cal I}_Q(m_Q)$, where all the ${\cal I}$'s approximately follow ${\cal I}_{\rho,\chi,Q} \propto e^{3.7m_Q}$.
For a given value of $g$, these terms would easily dominate \eqref{Friedmann},
\eqref{chiEoM} and \eqref{QEoM} for large $m_Q$, if one took $m_Q$ as a free parameter. However, this would infer that strong backreaction prevents the system from reaching such a parameter region. The conditions to ensure that each of $\rho_t$ and ${\cal T}_{BR}^{\chi,Q}$ is subdominant in \eqref{Friedmann}, \eqref{chiEoM} and \eqref{QEoM} are translated into upper bounds on $g$,
\begin{equation}
g < {\cal G}_{\rho,\chi,Q} (m_Q) ,
\end{equation}
where ${\cal G}_{\rho,\chi,Q} \propto {\cal I}_{\rho,\chi,Q}^{-1/2}$.
In Fig.~\ref{fig:checks}, we show the strongest constraints coming from $\mathcal{G}_\chi$, though they are almost degenerate.

\begin{figure*}
        \centering
        \begin{tabular}{cc}
                \begin{minipage}{0.5\hsize}
                        \centering
                        \vspace{11pt}
                        \includegraphics[width=\hsize]{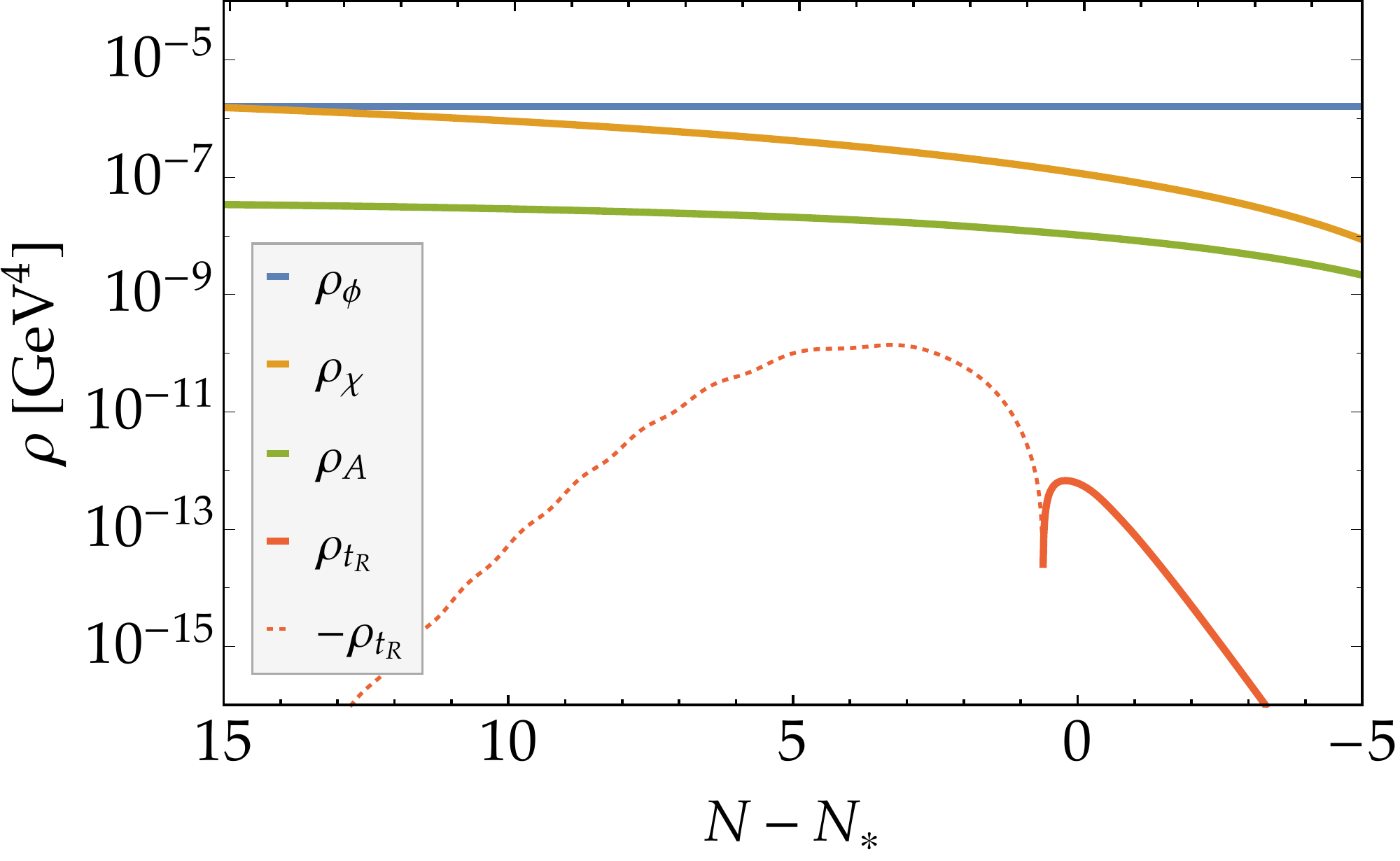}
                \end{minipage}
                \begin{minipage}{0.5\hsize}
                        \centering
                        \includegraphics[width=\hsize]{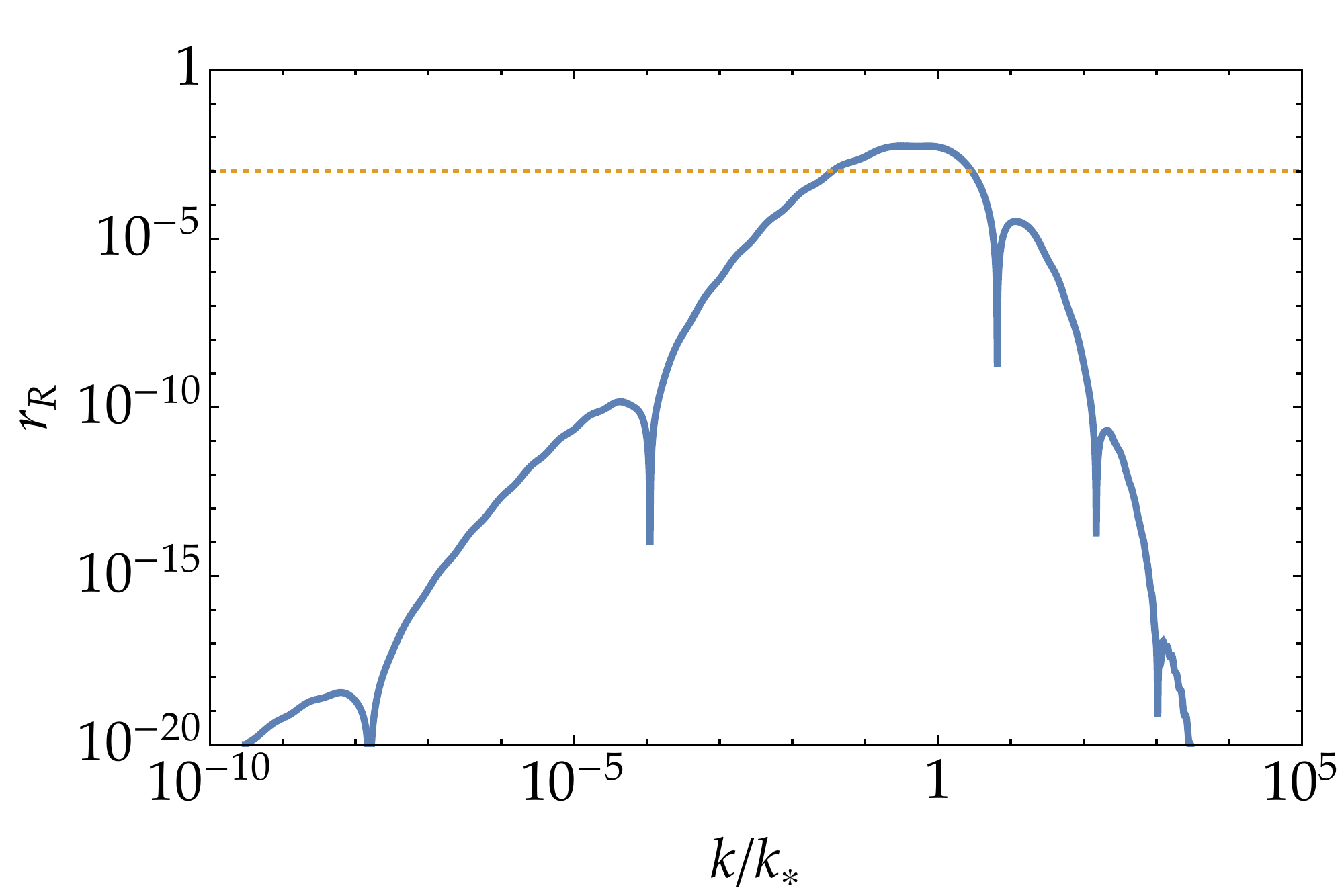}
                \end{minipage}
        \end{tabular}
        \caption{({\bf Left panel}) The energy densities of the inflaton $\rho_\phi$ (blue),
        the axion $\rho_\chi$ (yellow), the $SU(2)$ gauge field $\rho_A$ (green)
        and the backreaction from the amplified perturbation  $\rho_{t_R}$ (red) are shown. 
        The inflationary energy scale is as low as $\rho_\phi\approx (30{\rm MeV})^4$. 
        The horizontal axis represents the backward e-folds, and $N_*$ corresponds to the scale at which the 
        tensor perturbation is maximal. This scale can be identified as the CMB pivot scale since the inflaton sector is independent 
        of the gauge field sector. 
        Although $\rho_{t_R}$ is negative while the instability is getting stronger, the total energy of the $SU(2)$ gauge field is always positive.
        ({\bf Right panel}) The tensor-to-scalar ratio $r$ due to the sourced GW only with the right-handed
        polarization. On the CMB scale $k_*$ 
        it exceeds the threshold value $r=10^{-3}$ (yellow dashed line) and thus it is detectable for the upcoming CMB missions.}
        \label{fig: numerics}
\end{figure*}

For large $m_Q$, the backreaction is not completely negligible even in the allowed region shown in Fig.~\ref{fig:checks}.
In those cases, one has to resort to full numerical calculations simultaneously solving all equations of motion for 
background fields,~(\ref{Friedmann2})--(\ref{QEoM}) and for perturbations,~(\ref{tEoM}) and (\ref{psiEoM}) with full source terms included.
Fig.~\ref{fig: numerics} shows our numerical result for the following parameters:
\bae{\label{eq: paras}
        H_\text{inf}&=3\times10^{-22}\,\mathrm{GeV}, \quad \mu=0.055\,\mathrm{GeV}, \nonumber \\
        f&=1.5\times10^{17}\,\mathrm{GeV}, \quad \lambda=3000, \quad g=1.9\times10^{-36},
}
where the corresponding maximum of $m_Q$ is around 44.
The tensor-to-scalar ratio $r_R=\calP_{h_R}/\calP_\zeta(k_*)$ where $k_*$ is the pivot scale for CMB observations
indeed exceeds the detectable limit $10^{-3}$ even with such a extremely low inflationary energy scale $\sim 36 \, {\rm MeV}$.
%
%
%///////////////////////////////////////////////////////////////////////////////////%

%====================================================================================%
\subsection{B. Curvature Perturbation}
%====================================================================================%

Previous attempts to generate GW from scalar or vector fields are tightly constrained by the CMB observation on the curvature perturbation $\zeta$ 
\cite{Cook:2011hg, Ferreira:2014zia, *Ferreira:2015omg}. 
In our model, the inflaton fluctuation $\delta\phi$ is assumed to be responsible for generating $\zeta$ compatible with the CMB observation. 
Contributions from the other scalar modes $\delta\chi$, $\delta Q$ and $M$ to $\zeta$ are negligible, unless $\chi$ becomes a curvaton \cite{Dimastrogiovanni:2016fuu}.

In addition, we investigate another channel in which the second order effect of $t_R$ produces the inflaton perturbation, $t_R t_R \rightarrow \delta\phi$, through the gravitational interaction. This effect arises only at the second order due to the absence of linear couplings between $\delta\phi$ and $t_R$, while the sourcing of $t_R$ to the GW is first-order, thus $\zeta^{(s)} = - H \delta\phi^{(s)} / \dot{\phi} \propto (t_R / \Mpl)^2$ is expected to be negligible for the parameter range of our interest. We will address this effect in detail in the upcoming work.

Even though the part of $\delta\phi$ sourced by the second order of $t_R$ or the linear order of the scalar perturbations in the axion-$SU(2)$ sector only has negligible effects, that of $\delta\phi$ originated from its own vacuum fluctuations can be influenced by the background fields $\chi$ and $Q$, due to their contribution to $\dot{H}$.
As a result, the spectral index in our model reads,  
\begin{equation}
n_s-1= 2 \left( \eta_\phi - 3 \epsilon_\phi - \epsilon_\chi - \epsilon_A \right) 
\simeq 2 \left( \eta_\phi - \epsilon_B \right) ,
\end{equation}
where in the last step we have used $\epsilon_A \simeq \epsilon_B \gg \epsilon_\phi , \, \epsilon_\chi$, true with the parameters of our interest.
The Planck measures $n_s =0.9645 \pm 0.0049$ \cite{Ade:2015lrj}, and without assuming an accidental cancellation between $\epsilon_B$ and $\eta_\phi$, we require a bound on $\epsilon_B$ as
\begin{equation}
\epsilon_B(t_*) \lesssim 2\times 10^{-2},
\label{eB bound}
\end{equation}
where $t_*$ denotes the time of the horizon crossing of the CMB modes. Note that this constraint can be relaxed if $\eta_\phi$ is positive.
When \eqref{eB bound} saturates, in our model, $\epsilon_B$ can explain the red-tilted curvature perturbations without 
a huge hierarchy of slow-roll parameters $\eta_\phi\gg\epsilon_\phi$. 
It is a quite intriguing possibility for small-field inflationary models since all slow-roll parameters are naively expected to be equivalently small 
in that class of inflation. We numerically checked that $\ns(k_*)$ within 2$\sigma$ of Planck constraints is realized solely by $\epsilon_B$
for the parameters~(\ref{eq: paras}).

The bound \eqref{eB bound} is translated into a lower bound on $g= H m_Q^2/ (\Mpl \sqrt{\epsilon_B})$, with \eqref{Phs},
\begin{equation}
g= \frac{\pi m_Q^2}{\epsilon_B \mathcal{F}} \sqrt{r \mcP_\zeta}
\,\gtrsim\, 5\times 10^{-4} \frac{m_Q^2}{\mathcal{F}}
\left( \frac{r}{0.001} \right)^{1/2}
.
\end{equation}
We plot this as the light-blue shaded region in Fig.~\ref{fig:checks}.

%====================================================================================%
\subsection{C. Perturbativity}
%====================================================================================%

Since the amplitude of $t_R$ is substantially amplified due to the instability in our model, we need to ensure that it does not invalidate our perturbative calculation.
We thus impose that the $1$-loop contribution to the two-point function $\langle t_R t_R\rangle$ should be negligible to that of the tree level.
The terms $-F^a_{\mu\nu}F^{a\mu\nu}/4 + \lambda \chi F_{\mu\nu}^a \tilde{F}^{a\mu\nu}/(4f)$ lead to three- and four-point vertices, and it can be shown that their one-loop diagrams give contributions of the same order~\cite{Ferreira:2014zia}.
We here focus on the latter and demonstrate that the perturbativity condition gives no additional bounds on the model parameters. The four-point interaction Hamiltonian reads
\begin{align}
\hat{H}^{(4)}_I(\eta) &= \frac{g^2}{4} \int \dd^3x 
\left[ (\hat{t}_{ij} \hat{t}_{ij})^2 - \hat{t}_{ij} \hat{t}_{jl} \hat{t}_{lm} \hat{t}_{mi}\right],
\label{I Hamiltonian}
\end{align}
giving rise to, using the in-in formalism,
\begin{align}
&\langle \hat{t}_{ij}(\tau,\bm{k}) \hat{t}_{ij}(\tau,\bm{k}')\rangle_{\rm 1loop}
=\delta(\bm k + \bm k') \,
\frac{7g^2}{10\pi^2}  
\notag\\
& \times \int^\tau_{-\infty} \dd \eta\, {\rm Im}[t_R^2(k,\tau) t_R^{* \, 2}(k,\eta)] 
\int \dd \tilde k\, {\tilde k}^2\, |t_R(\tilde{k},\eta)|^2,
\label{1loop effect}
\end{align}
where we ignored the left-handed mode.
Defining ${\cal R}_t$ as the ratio of \eqref{1loop effect} divided by the tree-level contribution, $(2\pi)^3 \delta(\bm k + \bm k') |t_k(\tau)|^2$, evaluated at the time when $t_R$ reaches its maximum value, we ensure ${\cal R}_t \ll 1$ to safely ignore the higher-order loops and to justify the perturbative approach.
Evaluating \eqref{1loop effect} with \eqref{homogeneous solution}, 
we verify that $\mathcal{R}_t\ll1$ is satisfied up to $m_Q=50$ for $r=10^{-3}$.

%====================================================================================%
\section{IV. Conclusion}
%====================================================================================%

The main message of this Letter is that the detection of primordial gravitational waves does not necessarily exclude low-energy inflation.
Once an $SU(2)$ gauge field has a background configuration that respects the spatial rotation, its perturbations are coupled to the GW at the linear order. The former is amplified by instabilities around the horizon crossing, whose power is then linearly transferred to the latter. We have demonstrated that the GW power spectrum produced from this mechanism can be as significant as at detectable levels respecting all the consistency conditions, even if the inflationary energy scale is close to the BBN bound.

Having a possible alternative source of GW, it is crucial to discriminate the generation mechanism of primordial GW to reveal the true energy scale of inflation. Fortunately, our model has the following distinct predictions
to be distinguished from the conventional vacuum GW.
(i) The fully parity-violating GW may be detected through 
CMB temperature and B-mode (TB) or E-mode and B-mode polarization
(EB) cross-correlation by the upcoming satellite mission such as LiteBIRD~\cite{Thorne:prep}.
(ii) Our model produces a sizable tensor non-Gaussianity with a particular
shape~\cite{Agrawal:prep}. (iii) The conventional consistency relation, $n_T=-r_{\rm vac}/8$, is broken, where $n_T$ is the tensor spectral index.
With the future observation, these signatures will carry important information for rigorous determination of inflationary energy scale.

\section{Acknowledgement}

We would like to thank Emanuela Dimastrogiovanni, Matteo Fasiello, Shinji Mukohyama, Marco Peloso, Matthew Reece, Martin Sloth, Henry Tye and Yi Wang for useful discussions and correspondences.
TF acknowledges the support by Grant-in-Aid for JSPS Fellows No.~29-9103. 
RN is supported by the Natural Sciences and Engineering Research Council (NSERC) of Canada and by the Lorne Trottier Chair in Astrophysics and Cosmology at McGill University.
YT is supported by Japan Society for the Promotion of Science Research Fellowship for Young Scientists
and grants from Région Île-de-France.

%====================================================================================%
%  References
%====================================================================================%

\bibliography{FNT_letter}

%%%%%%%%%%%%%%%%%%%%%%%%%%%%%%%%%%%%%%%
\begin{comment}

\end{comment}
%%%%%%%%%%%%%%%%%%%%%%%%%%%%%%%%%%%%%

\end{document}